\documentclass[a4paper, 11pt]{article}

\usepackage[utf8]{inputenc}
\usepackage[T1]{fontenc}

\usepackage[ttscale=.875]{libertine}
\usepackage[libertine]{newtxmath}

\usepackage[margin=2.5cm]{geometry}
\usepackage{booktabs,dcolumn,rotating}
\usepackage{microtype}
\usepackage{color}
\usepackage{amsfonts}

\begin{document}

\title{\textbf{Secure by default -- the case of TLS}}

\author{Martin Stanek \\[1ex]
  {\small Department of Computer Science} \\ 
  {\small Comenius University} \\
  {\small\mbox{stanek@dcs.fmph.uniba.sk}}}
\date{}
\maketitle

\begin{abstract}
Default configuration of various software applications often neglects security objectives.
We tested the default configuration of TLS in dozen web and application servers. 
The results show that ``secure by default'' principle should be adopted more broadly 
by developers and package maintainers. In addition, system administrators cannot  
rely blindly on default security options. 
\end{abstract}

{\small\textbf{Keywords:} TLS, secure defaults, testing.}

\section{Introduction}

Security often depends on prudent configuration of software components used in a deployed
system. All necessary security controls and options are there, but one have 
to turn them on or simply start using them. Unfortunately, the ``If it ain’t broke, 
don’t fix it'' philosophy or a lack of expertise wins sometimes. The technology is 
deployed in a default configuration or configuration that fulfills (mostly functional)
requirements with as few changes as possible. 

Secure by default is a well known security principle, see e.g. \cite{NCSC}:

\begin{quote}
\emph{Technology which is Secure by Default has the best security it can without you even knowing 
it's there, or having to turn it on.}
\end{quote}

\noindent We should aim to provide software packages with safe defaults and turning them to less
secure configuration should require a deliberate effort, see e.g. \cite{OWASP}:

\begin{quote}
\emph{There are many ways to deliver an ``out of the box'' experience for users. However, by
default, the experience should be secure, and it should be up to the user to reduce 
their security -- if they are allowed.}
\end{quote}

The Transport Layer Security (TLS) and its predecessor Secure Socket Layer (SSL) are 
widely used protocols for ensuring confidentiality and integrity of transported data, 
as well as one or two-sided authentication of communicating parties. There are various 
versions of the protocols and extensions proposed and implemented during more 
than 20 years of using SSL/TLS. Meanwhile, many flaws, both theoretical and practical 
were found in the design and implementation of the protocols \cite{MS13,SHS15}. 
New weaknesses are constantly found; in recent years we have seen DROWN \cite{DROWN}
or LOGJAM \cite{LOGJAM}.
It is not easy to stay focused, understand the flaws and mitigate them in a timely 
fashion. Moreover, the compatibility with client software (web browsers) should 
be considered.

We tested dozen web and application servers for default SSL/TLS configuration 
to see how the principle of secure defaults is applied in this context. We used 
the default configuration of the servers, and we made minimal changes to enable 
SSL/TLS, usually pointing to private key and public certificate 
file(s). We used an excellent tool testssl.sh by Dirk Wetter \cite{Wetter} designed 
for testing servers for security weaknesses in their SSL/TLS configuration.

The results are not surprising. We observed diverse combinations of default settings.
The ``secure by default'' principle is followed rarely. 
Detailed results are presented in Section \ref{sec-result}. Let us emphasize that 
different versions of the servers or even the same versions packaged for another operating 
system or distribution can use different defaults, thus leading to different results.

\section{Test environment}

The test environment was based on Ubuntu 16.04 LTS distribution with up-to-date updates. 
We chose popular, well-known web and applications servers as well as some small
players in this area.  
Some servers were installed from available Ubuntu repositories, some were downloaded as 
binaries from projects' web sites, some were compiled from source code, and puma 
(a Ruby web server) was installed via RubyGems. Table \ref{tab1} summarizes all twelve 
servers, their versions, and how they were obtained. In some cases, namely for gunicorn 
and puma, a minimal ``Hello world!'' application was created. 

\begin{table}[!ht]
\centering
\begin{tabular}{lrl}
  \toprule
  server & version & comment \\
  \midrule 
  apache & 2.4.18 &  package for Ubuntu 16.04 \\
  caddy & 0.10.4 & dowloaded from caddyserver.com \\
  glassfish & 4.1.2 & downloaded from javaee.github.io/glassfish \\
  gunicorn & 19.4.5 &  package for Ubuntu 16.04\\
  hiawatha & 10.6 & downloaded from www.hiawatha-webserver.org, compiled \\
  jetty & 9.2.14 &  package for Ubuntu 16.04 \\
  lighttpd & 1.4.35 &  package for Ubuntu 16.04 \\
  nginx & 1.10.3 &  package for Ubuntu 16.04 \\
  openlitespeed & 1.4.26 & downloaded from open.litespeedtech.com, compiled \\
  puma & 3.9.1 & installed via ``gem install puma'' \\
  tomcat & 8.0.32 &  package for Ubuntu 16.04 \\
  wildfly & 10.1.0 & downloaded from wildfly.org \\
  \bottomrule 
\end{tabular} 
\caption{Versions of tested servers}
\label{tab1}
\end{table}

Most of the servers allow fine-grained configuration of TLS support. We intentionally 
use the default configuration and made only minimal changes to enable TLS (following 
a documentation in this regard). It is not uncommon to see administrators applying 
such approach when managing IT systems. 
The tool testssl.sh \cite{Wetter} was used for comprehensive test of each server's 
TLS configuration.

\section{Results}
\label{sec-result}

\newcommand{\xmark}{$\times$}
\newcommand{\rxmark}{{\color{red} $\times$}}
\newcommand{\rcheckmark}{{\color{red} \checkmark}}

The results are presented in a series of tables with comments. We use marks \checkmark
and \xmark\ to indicate supported/unsupported feature or immunity/susceptibility to
a vulnerability. Moreover, we indicate by red color all those marks that we think the 
secure configuration should not have.

\subsection{Protocol versions}

There are several versions of SSL/TLS protocol. The good news is that both SSL 
versions (2.0 and 3.0) are disabled by default in all tested servers. The situation 
is more interesting for TLS versions. The majority of the servers, including 
``big names'' like apache, nginx, tomcat or wildfly allow all three TLS versions.
Two web servers caddy and hiawatha do not enable TLS 1.0 by default, which is 
a right thing to do. A specific case is gunicorn server --- it accepts only 
TLS 1.0 by default. Detailed information is shown in Table \ref{tab-versions}.

\begin{table}[!ht]
\centering
\begin{tabular}{lcccc}
  \toprule
  server & SSL ($\leq$ 3.0) & TLS 1.0 & TLS 1.1 & TLS 1.2 \\
  \midrule 
  apache & \xmark & \rcheckmark & \checkmark & \checkmark \\
  caddy  & \xmark & \xmark & \checkmark & \checkmark \\
  glassfish & \xmark & \rcheckmark & \checkmark & \checkmark \\
  gunicorn  & \xmark & \rcheckmark & \rxmark    & \rxmark \\
  hiawatha  & \xmark & \xmark & \checkmark & \checkmark \\
  jetty     & \xmark & \rcheckmark & \checkmark & \checkmark \\
  lighttpd  & \xmark & \rcheckmark & \checkmark & \checkmark \\
  nginx     & \xmark & \rcheckmark & \checkmark & \checkmark \\
  openlitespeed & \xmark & \rcheckmark & \checkmark & \checkmark \\
  puma  	 & \xmark & \rcheckmark & \checkmark & \checkmark \\
  tomcat  & \xmark & \rcheckmark & \checkmark & \checkmark \\
  wildfly & \xmark & \rcheckmark & \checkmark & \checkmark \\
  \bottomrule 
\end{tabular} 
\caption{Default support for SSL/TLS protocol versions}
\label{tab-versions}
\end{table}

\subsection{Ciphers}

Web servers should set a reasonable cipher order for supported ciphers. Some of the tested servers
do not define their cipher order at all (again, in their default configuration). Another serious 
weakness is using NULL cipher or obsolete export-grade ciphers --- this happens in one server (gunicorn).
Few servers offer weak 128-bit ciphers (RC4 or SEED). 
An interesting situation is the default support for TripleDES algorithm --- a clear consensus is 
missing. A prudent approach is to turn it off.

\begin{table}[!ht]
\centering
\begin{tabular}{lccccccc}
  \toprule
  server & Cipher order & N/A/E$^{(1)} $ & LOW$^{(2)}$ & Weak$^{(3)}$ & TripleDES & High$^{(4)}$ & AEAD$^{(5)}$ \\
  \midrule 
  apache    & \rxmark & \xmark/\xmark/\xmark & \xmark & \xmark & \xmark & \checkmark & \checkmark \\
  caddy     & \checkmark & \xmark/\xmark/\xmark & \xmark & \xmark & \xmark & \checkmark & \checkmark \\
  glassfish & \rxmark & \xmark/\xmark/\xmark & \xmark & \xmark & \rcheckmark & \checkmark & \checkmark \\
  gunicorn  & \rxmark & \rcheckmark/\rcheckmark/\xmark & \xmark & \rcheckmark & \rcheckmark & \checkmark & \rxmark \\
  hiawatha  & \checkmark & \xmark/\xmark/\xmark & \xmark & \xmark & \xmark & \checkmark & \checkmark \\
  jetty     & \rxmark & \xmark/\xmark/\xmark & \xmark & \xmark & \rcheckmark & \checkmark & \checkmark \\
  lighttpd  & \checkmark & \xmark/\xmark/\xmark & \xmark & \rcheckmark & \xmark & \checkmark & \rxmark \\
  nginx     & \checkmark & \xmark/\xmark/\xmark & \xmark & \xmark & \xmark & \checkmark & \checkmark \\
  openlitespeed & \checkmark & \xmark/\xmark/\xmark & \xmark & \xmark & \xmark & \checkmark & \checkmark \\
  puma      & \checkmark & \xmark/\xmark/\xmark & \xmark & \xmark & \xmark & \checkmark & \checkmark \\
  tomcat    & \rxmark & \xmark/\xmark/\xmark & \xmark & \xmark & \rcheckmark & \checkmark & \checkmark \\
  wildfly   & \rxmark & \xmark/\xmark/\xmark & \xmark & \xmark & \rcheckmark & \checkmark & \checkmark \\
  \bottomrule  \\[-1ex]
  \multicolumn{8}{l}{$^{(1)}$ NULL or Anonymous NULL or Export ciphers} \\
  \multicolumn{8}{l}{$^{(2)}$ LOW --- DES or $64$-bit ciphers} \\
  \multicolumn{8}{l}{$^{(3)}$ Weak $128$-bit ciphers like IDEA, SEED, RC4 etc.} \\
  \multicolumn{8}{l}{$^{(4)}$ Strong ciphers without AEAD} \\
  \multicolumn{8}{l}{$^{(5)}$ AEAD (Authenticated Encryption with Associated Data) ciphers} \\
\end{tabular} 
\caption{Categories of ciphers enabled/offered (indicated by \checkmark)}
\label{tab-ciphers}
\end{table}

A positive fact is that all servers support some cipher suites offering PFS (Perfect forward secrecy).

\subsection{Vulnerabilities}

Not all vulnerabilities are equally dangerous. For example, not implementing TLS Fallback
Signaling Cipher Suite Value or accepting ciphers in CBC mode (potentially vulnerable to 
LUCKY13, though the timing side channels are often already fixed in underlying TLS 
implementations) is less serious than accepting RC4 cipher or using common group 
of small order (LOGJAM).
In some cases, the web browser used by a user determines whether the weakness 
is a valid threat. For example, the BEAST attack is prevented in sufficiently recent 
browser versions (by 1/n-1 record splitting). SSL Labs does not even track this 
vulnerability in their SSL Pulse anymore
\cite{SSLPulse}.

Table \ref{tab-vulns} contains results for the following vulnerabilities: Heartbleed
(CVE-2014-0160), CCS (CVE-2014-0224), Secure Renegotiation (CVE-2009-3555), Secure 
Client-Initiated Renegotiation, CRIME (CVE-2012-4929), BREACH (CVE-2013-3587), 
POODLE (CVE-2014-3566), TLS FALLBACK SCSV (RFC 7507), SWEET32 (CVE-2016-2183, 
CVE-2016-6329), FREAK (CVE-2015-0204), DROWN (CVE-2016-0800, CVE-2016-0703),
LOGJAM (CVE-2015-4000), BEAST (CVE-2011-3389), LUCKY13 (CVE-2013-0169),
RC4 (CVE-2013-2566, CVE-2015-2808). We do not explain the vulnerabilities in detail, 
a curious reader can find additional information in corresponding CVEs, RFCs and 
references therein.

\newcolumntype{R}[1]{>{\begin{turn}{90}\begin{minipage}{#1}\normalsize}l%
<{\end{minipage}\end{turn}}%
}

\begin{table}[!ht]
\centering
\begin{tabular}{lccccccccccccccc}
  \toprule
  server & \multicolumn{1}{R{4cm}}{Heartbleed} 
         & \multicolumn{1}{R{4cm}}{CCS}
         & \multicolumn{1}{R{4cm}}{Secure Renegotiation}
         & \multicolumn{1}{R{4cm}}{Secure Client-Init. Reneg.}
         & \multicolumn{1}{R{4cm}}{CRIME} 
         & \multicolumn{1}{R{4cm}}{BREACH} 
         & \multicolumn{1}{R{4cm}}{POODLE} 
         & \multicolumn{1}{R{4cm}}{TLS FALLBACK SCSV} 
         & \multicolumn{1}{R{4cm}}{SWEET32} 
         & \multicolumn{1}{R{4cm}}{FREAK} 
         & \multicolumn{1}{R{4cm}}{DROWN} 
         & \multicolumn{1}{R{4cm}}{LOGJAM} 
         & \multicolumn{1}{R{4cm}}{BEAST} 
         & \multicolumn{1}{R{4cm}}{LUCKY13$^{(\text{c})}$} 
         & \multicolumn{1}{R{4cm}}{RC4} \\
  \midrule 
  apache     & \checkmark & \checkmark & \checkmark & \checkmark & \checkmark & \rxmark$^{(\text{a})}$ & \checkmark & \checkmark 
             & \checkmark & \checkmark & \checkmark & \rxmark$^{\text{(b-14)}}$ & \rxmark & \xmark & \checkmark \\
  caddy      & \checkmark & \checkmark & \checkmark & \checkmark & \checkmark & \checkmark & \checkmark & \checkmark 
             & \checkmark & \checkmark & \checkmark & \checkmark & \checkmark & \xmark & \checkmark \\
  glassfish  & \checkmark & \checkmark & \checkmark & \rxmark & \checkmark & \checkmark & \checkmark & \rxmark 
             & \rxmark & \checkmark & \checkmark & \rxmark$^{\text{(b-2)}}$ & \rxmark & \xmark & \checkmark \\
  gunicorn   & \checkmark & \checkmark & \checkmark & \rxmark & \checkmark & \checkmark & \checkmark & \rxmark 
             & \rxmark & \checkmark & \checkmark & \checkmark & \rxmark & \xmark & \rxmark \\
  hiawatha   & \checkmark & \checkmark & \checkmark & \checkmark & \checkmark & \rxmark$^{(\text{a})}$ & \checkmark & \checkmark 
             & \checkmark & \checkmark & \checkmark & \rxmark$^{\text{(b-r)}}$ & \checkmark & \xmark & \checkmark \\
  jetty      & \checkmark & \checkmark & \checkmark & \rxmark & \checkmark & \checkmark & \checkmark & \rxmark 
             & \rxmark & \checkmark & \checkmark & \rxmark$^{\text{(b-2)}}$ & \rxmark & \xmark & \checkmark \\
  lighttpd   & \checkmark & \checkmark & \checkmark & \checkmark & \checkmark & \rxmark$^{(\text{a})}$ & \checkmark & \checkmark 
             & \checkmark & \checkmark & \checkmark & \checkmark & \rxmark & \xmark & \rxmark \\
  nginx      & \checkmark & \checkmark & \checkmark & \checkmark & \checkmark & \rxmark$^{(\text{a})}$ & \checkmark & \checkmark 
             & \checkmark & \checkmark & \checkmark & \rxmark$^{\text{(b-n)}}$ & \rxmark & \xmark & \checkmark \\
  openlitespeed   & \checkmark & \checkmark & \checkmark & \checkmark & \checkmark & \rxmark$^{(\text{a})}$ & \checkmark 
             & \checkmark  & \checkmark & \checkmark & \checkmark & \checkmark & \rxmark & \xmark & \checkmark \\
  puma       & \checkmark & \checkmark & \checkmark & \rxmark & \checkmark & \checkmark & \checkmark & \checkmark 
             & \checkmark & \checkmark & \checkmark & \checkmark & \rxmark & \xmark & \checkmark \\
  tomcat     & \checkmark & \checkmark & \checkmark & \checkmark & \checkmark & \checkmark & \checkmark & \rxmark 
             & \rxmark & \checkmark & \checkmark & \rxmark$^{\text{(b-2)}}$ & \rxmark & \xmark & \checkmark \\
  wildfly    & \checkmark & \checkmark & \checkmark & \rxmark & \checkmark & \checkmark & \checkmark & \rxmark 
             & \rxmark & \checkmark & \checkmark & \rxmark$^{\text{(b-2)}}$ & \rxmark & \xmark & \checkmark \\
  \bottomrule \\[-1ex]
  \multicolumn{16}{l}{$^{(\text{a})}$ gzip compression of ``/'' path} \\
  \multicolumn{16}{l}{$^{(\text{b-2})}$ no DH EXPORT but common group is used (Oakley Group 2, 1024-bit)} \\
  \multicolumn{16}{l}{$^{(\text{b-14})}$ no DH EXPORT but common group is used (Oakley Group 14, 2048-bit)} \\
  \multicolumn{16}{l}{$^{(\text{b-r})}$ no DH EXPORT but common group is used (RFC5114/2048-bit group)} \\
  \multicolumn{16}{l}{$^{(\text{b-n})}$ no DH EXPORT but common group is used (nginx/1024-bit group)} \\
  \multicolumn{16}{l}{$^{(\text{c})}$ indicates ciphers in CBC mode (not actual existence of timing side channels)}   
\end{tabular} 
\caption{Default configuration potentially vulnerable (\xmark) or not (\checkmark)}
\label{tab-vulns}
\end{table}

\section{Conclusion}

We tested dozen web and application servers for security of their default TLS configuration.
The results show that ``secure by default'' principle is seldom followed in this, and probably
other areas as well.
The system administrators cannot rely blindly on default security options. This increases 
the (hopefully obvious) importance of thorough review and setting of security configuration.


\end{document}